\documentclass[useAMS,usenatbib]{mn2e}
\usepackage{graphicx}
\usepackage{amssymb}
\usepackage[modulo,switch]{lineno}
\title[Frequency dependence of $\Delta \nu$: Influence of the He II zone]
  {Frequency dependence of the large frequency separation of solar-like oscillators: Influence of the Helium second-ionization zone}
\author[S. Hekker et al.]
  {S.~Hekker$^{1,2}$, Sarbani Basu$^3$, Y. Elsworth$^2$, W.J. Chaplin$^2$\\
  $^1$ Astronomical Institute `Anton Pannekoek', University of Amsterdam, Science Park 904, 1098 HX Amsterdam, the Netherlands\\
  $^2$ School of Physics and Astronomy, University of Birmingham, Edgbaston, Birmingham B15 2TT, UK\\
  $^3$ Department of Astronomy, Yale University, P.O. Box 208101, New Haven CT 06520-8101, USA}

\pagerange{\pageref{firstpage}--\pageref{lastpage}} \pubyear{2011}

\def\LaTeX{L\kern-.36em\raise.3ex\hbox{a}\kern-.15em
    T\kern-.1667em\lower.7ex\hbox{E}\kern-.125emX}

\begin{document}
\linenumbers
\newcommand{\meandnu} {\langle\Delta\nu\rangle}
\label{firstpage}

\maketitle

\begin{abstract}
The large frequency separation ($\Delta \nu$) between modes of the same degree and consecutive orders in a star is approximately proportional to the square root of its mean density. To determine $\Delta \nu$ as accurately as possible a mean large frequency separation ($\meandnu$) computed over several orders is often used. It is, however, known that $\Delta \nu$ varies with frequency in a second order effect. From observations it has been shown that this frequency dependence is more important for main-sequence stars than it is for red-giant stars. 
Here we use YREC models to verify and explain this observational result. 
We find that for stars with $R \gtrsim 8$~R$_{\odot}$ the effect of the Helium second ionisation zone is relatively small. For these stars the deep location of the He II zone induces a frequency modulation covering only a few $\Delta \nu$, while the amplitude of the modulation is low due to the relatively weak and extended He II layer, causing a shallow wide depression in $\Gamma_1$. For less evolved stars the He II zone is located closer to the surface, and it is more confined, i.e. a deep narrow depression in $\Gamma_1$. This causes frequency modulations with relatively high amplitudes covering up to about 20$\Delta \nu$, inducing a relatively large frequency modulation. Additionally, we find that for less evolved stars the He II zone is stronger and more localised for more massive stars and for stars with low metallicities further increasing the amplitude of the frequency modulation. 
\end{abstract}


\section{Introduction}
Stellar oscillations can be used to determine the internal structures of stars. Solar-type stars, subgiants and red-giant stars exhibit solar-like oscillations, i.e., oscillations stochastically excited in the turbulent outer layer. These oscillations generally appear in regular patterns in frequency following to reasonable approximation the asymptotic relation derived by \citet{tassoul1980}:
\begin{equation}
\nu_{n,l} \approx \Delta\nu(n+\ell/2+\epsilon)-\ell(\ell+1)D_0,
\label{tassoul}
\end{equation}
with $\nu_{n,l}$ the frequency of an oscillation mode with radial order $n$ and degree $\ell$ and $\Delta \nu$ the large frequency separation between modes of the same degree and consecutive orders. $D_0$ is most sensitive to deeper layers in the star and $\epsilon$ to the surface layers.

\begin{figure}
\begin{minipage}{\linewidth}
\centering
\includegraphics[width=\linewidth]{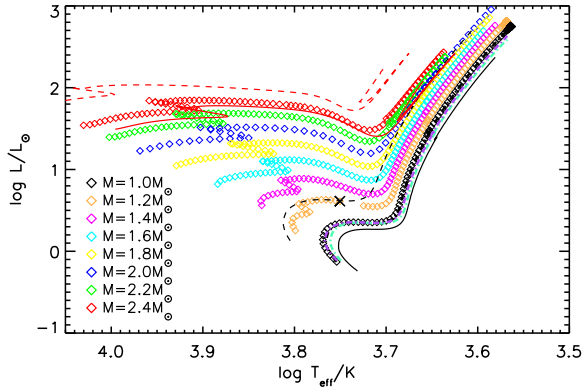}
\end{minipage}
\caption{YREC evolutionary tracks in a Hertzsprung-Russell diagram. Tracks with approximately solar metallicity are indicated with diamonds, while the solid lines show the tracks with [Fe/H] = +0.3 and the dashed lines indicate the tracks with [Fe/H] = $-$0.5 for models with $M=1.0$~M$_{\odot}$ (black) and 2.4 M$_{\odot}$ (red). Models with $M=1.0$~M$_{\odot}$, [Fe/H] = 0.0 including either diffusion, Krishna Swamy $T$-$\tau$ relation or $\alpha=1.75~H_p$ (see text for more details) are indicated with the black, mint and purple dashed-dotted lines, respectively. These lines nearly coincide with the solar metallicity $M=1.0$~M$_{\odot}$ models indicated with the black diamonds. The black cross indicates the model used in Fig.~\ref{dnunu}.}
\label{hrd}
\end{figure}

\begin{figure}
\begin{minipage}{\linewidth}
\centering
\includegraphics[width=\linewidth]{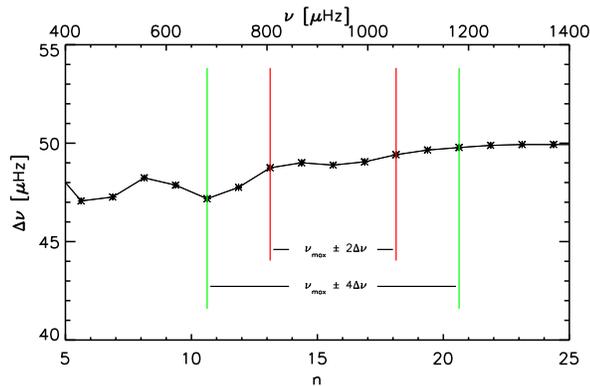}
\end{minipage}
\caption{$\Delta \nu$ computed using individual frequencies of $\ell=0$ modes as a function of radial order (bottom axis) or frequency (top axis) for the 1.2~M$_{\odot}$ star indicated with the black cross in Fig.~\ref{hrd}. The intervals used to compute $\meandnu$ are indicated with the red and green vertical lines.}
\label{dnunu}
\end{figure}

The large frequency separation is inversely proportional to the sound travel time through the star and it can be shown that this is directly proportional to the square root of the mean density of the star \citep{ulrich1986,kjeldsen1995}. To a first order approximation the large frequency separation is constant over the observed frequency range and often a mean large frequency separation ($\meandnu$) is computed. However, for the Sun and other main-sequence stars it has been shown that the frequency range over which $\meandnu$ is computed has a significant influence on the resulting value. This frequency dependence can be attributed to acoustic glitches, i.e., sudden internal property changes at the base of the convection zone and at the Helium second-ionization zone. Also other more slowly varying underlying variations, such as changing conditions close to the core, can cause departures from uniform $\Delta \nu$. Theory and methods considering acoustic glitches were first explored in the context of studies of solar oscillations, as initially proposed in \citet{gough1990}, and developed further by e.g. \citet{basu1994}, \citet{monteiro1994}, \citet{perez1998}, \citet{basu2004}. Different methods of analysis for extracting data on the glitches have been developed \citep[e.g. see][]{mazumdar2001a,mazumdar2001b,ballot2004,verner2006, houdek2007,roxburgh2010}, in some cases for application to other stars. Recently, the HeII ionization zone glitch signal has been measured in the red giant HR7349 \citep{carrier2010,miglio2010}. Additionally, \citet{mazumdar2010} have claimed detection of acoustic glitches, from the HeII ionization zone and the base of the convective envelope, in solar-like oscillations of the F-type star HD49933.


In contrast to the frequency dependence of $\meandnu$ for the Sun and main-sequence stars, first results from the \textit{Kepler} mission \citep{borucki2010} showed that for red giants the value of $\meandnu$ is much less sensitive to the frequency range over which it was computed \citep{hekker2011comp}. Here we investigate the reason for the difference in sensitivity of $\meandnu$ to the frequency range for main-sequence stars and red giants using models constructed using YREC, the Yale stellar evolution code \citep{demarque2008}.

\section{YREC Models}
A full overview of the YREC code can be found in \citet{demarque2008}. We use OPAL opacities \citep{iglesias1996} supplemented with low temperature ($\log T < 4.1$) opacities of \citet{ferguson2005} and the OPAL equation of state \citep{rogers2002}. All nuclear reaction rates are obtained from \citet{adelberger1998}, except for that of the $^{14}N(p,\gamma)^{15}O$ reaction, for which we use the rate of \citet{formicola2004}. 

In this work we use a sequence of  models of solar metallicity with masses between 1.0 and 2.4 M$_{\odot}$. The sequences range from just after ZAMS to near the tip of the giant branch for stars of masses up to 2.0 M$_{\odot}$ and up to the helium burning stage for the higher mass models. Additionally we also use 1.0 and 2.4 M$_{\odot}$ models with [Fe/H]=+0.3 and $-0.5$ to investigate the effects of metallicity. The masses and metallicities are chosen to bracket the masses and metallicities of stars expected to be observed with \textit{Kepler}. The models were constructed without diffusion and gravitational settling of helium and heavy elements. However, we do examine the role of diffusion and gravitational settling using the \citet{thoul1994} prescription for a sequence of 1.0 M$_{\odot}$ models. All models described here were constructed to have the Eddington $T$-$\tau$ relation in their atmosphere.
To study the effect of atmospheric structure we use a sequence of 1.0 M$_{\odot}$ models constructed with the Krishna Swamy $T$-$\tau$ relation \citep{krishnaswamy1966}. Additionally, to examine the effect of the mixing-length parameter, $\alpha$, we have made a sequence of 1.0 M$_{\odot}$ models with $\alpha=1.75~H_p$ instead of the solar value of $\alpha=1.826~H_p$ used in the other models.  Evolutionary tracks of the models are shown in Fig.~\ref{hrd}.

Note that we consider solar-like oscillations in models of main-sequence stars up to masses of 2.4 M$_{\odot}$, although solar-like oscillations are generally not observed in main-sequence stars with masses above roughly 1.5 M$_{\odot}$. Furthermore, we investigate models with radii less than 20~R$_{\odot}$ and with $\Delta\nu > 2~\mu$Hz compatible with the observational results by \citet{hekker2011comp}.

\begin{figure*}
\begin{minipage}{0.45\linewidth}
\centering
\includegraphics[width=\linewidth]{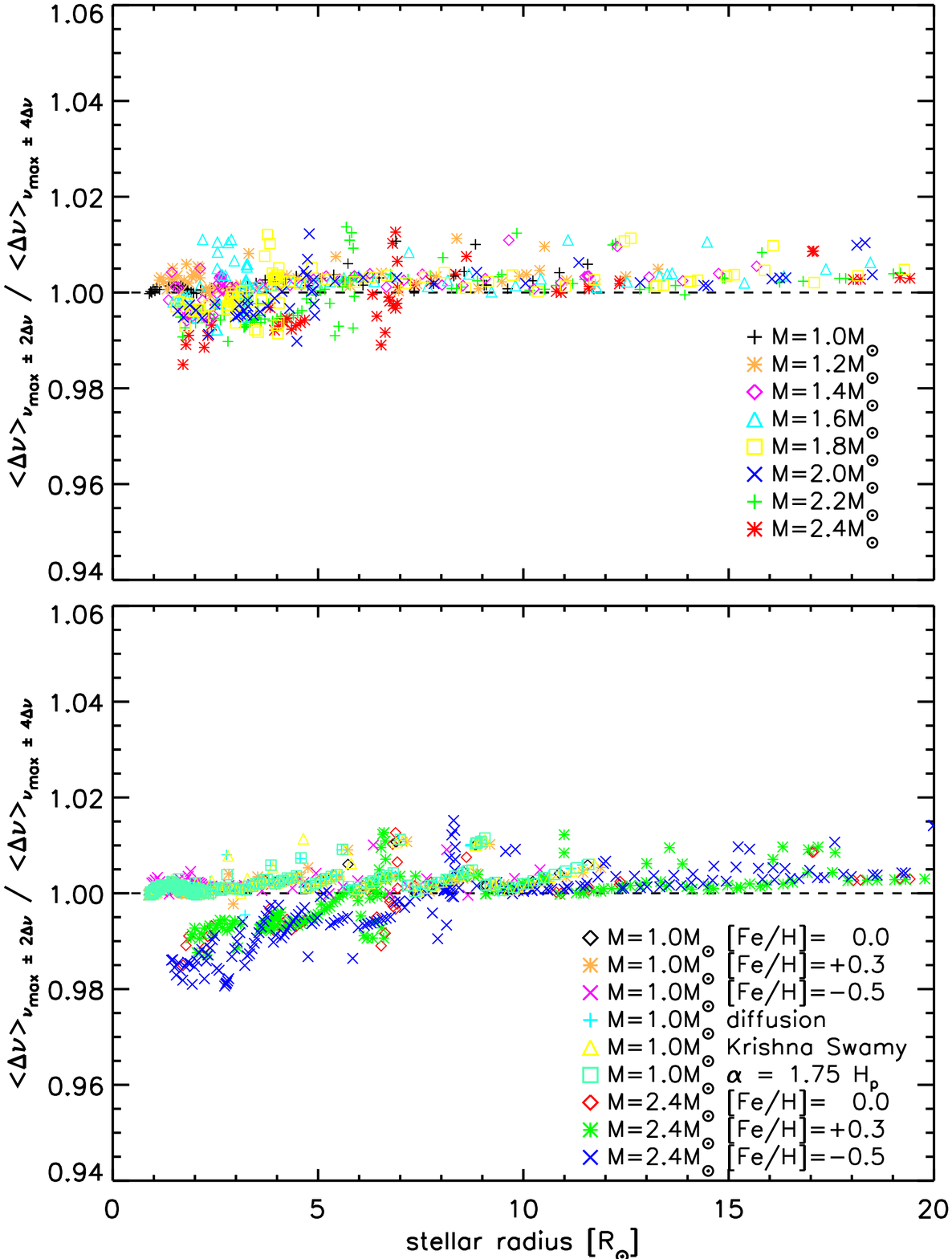}
\end{minipage}
\begin{minipage}{0.45\linewidth}
\centering
\includegraphics[width=\linewidth]{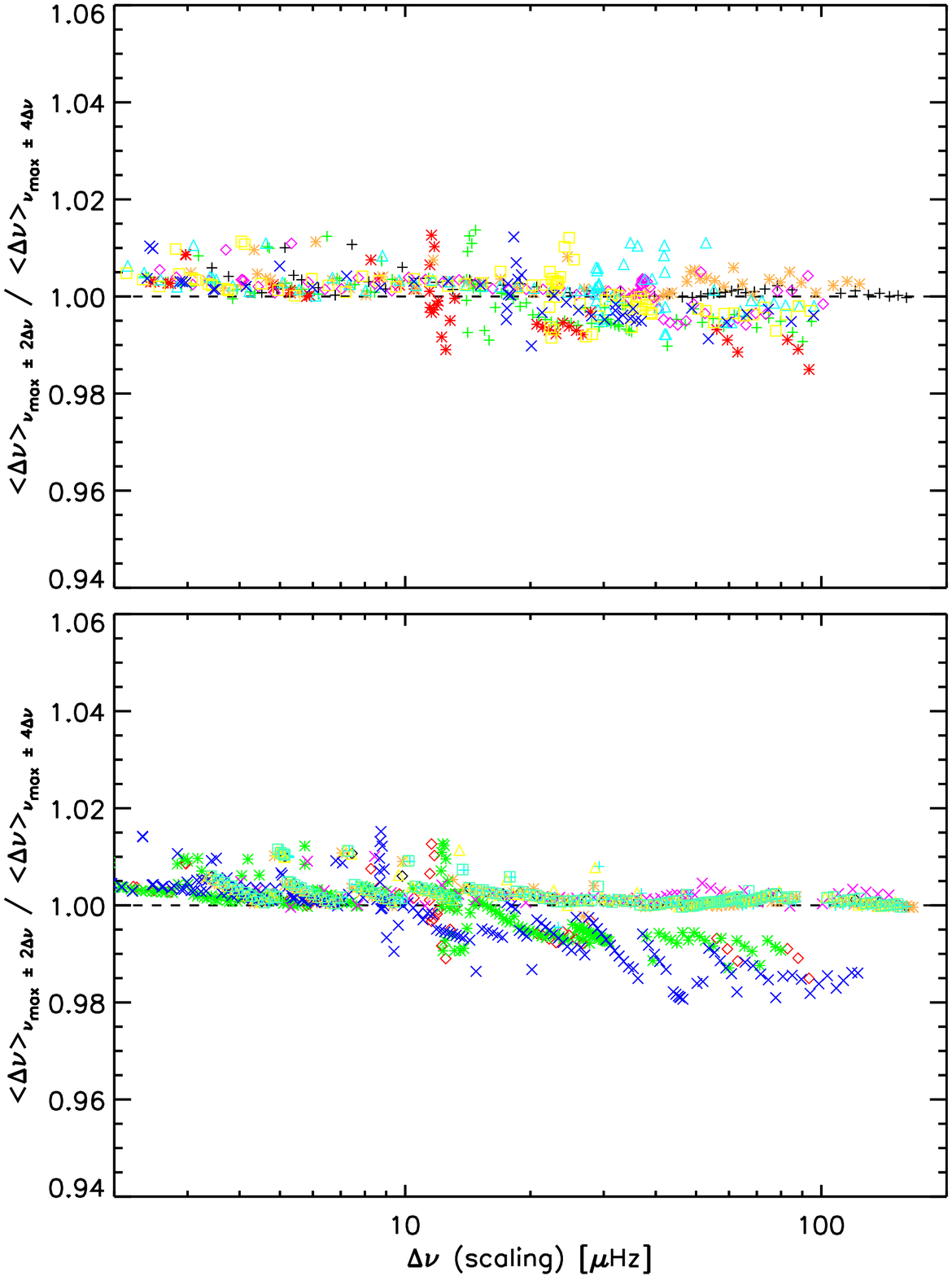}
\end{minipage}
\caption{Left: Ratio of $\meandnu$ values computed from a linear fit to the frequencies versus the radial order of the modes in the intervals $\nu_{\rm max} \pm 2\Delta \nu$ and $\nu_{\rm max} \pm 4\Delta \nu$ as a function of stellar radius for different masses (top) and different metallicities  or diffusion (bottom). Right: Same as right panels but now as a as a function of $\meandnu$. The dashed black line indicates unity.}
\label{ratiofit}
\end{figure*}

\section{Dependence of $\meandnu$ on frequency range}
To investigate the frequency dependence of $\meandnu$ from the models, we derived $\meandnu$ from linear fits of the frequencies of radial modes in the intervals $\nu_{\rm max} \pm 2(\Delta \nu)_{\rm scaling}$ and  $\nu_{\rm max} \pm 4(\Delta \nu)_{\rm scaling}$ as a function of their radial orders (see Fig.~\ref{dnunu} and Eq.~\ref{tassoul}). $\nu_{\rm max}$ is the frequency of maximum oscillation power estimated from the mass, radius and effective temperature of the models \citep{brown1991} and $(\Delta \nu)_{\rm scaling}$ is the mean large separation obtained from the mass and radius of the star, using scaling relations \citep{ulrich1986,kjeldsen1995}. We also investigated the median $\Delta \nu$ values in the respective frequency intervals to investigate the symmetry of $\Delta\nu$ in frequency. This showed that the behaviour of $\Delta \nu$ with frequency is antisymmetric about $\nu_{\rm max}$ and that a linear fit can be used in a first order approximation. We note that in those cases where it is not possible to extract individual radial-mode frequencies from real data, techniques are usually applied to the full oscillation power spectrum to estimate $\meandnu$, in which case the estimated value is also affected by non-radial modes.

\begin{figure*}
\begin{minipage}{0.45\linewidth}
\centering
\includegraphics[width=\linewidth]{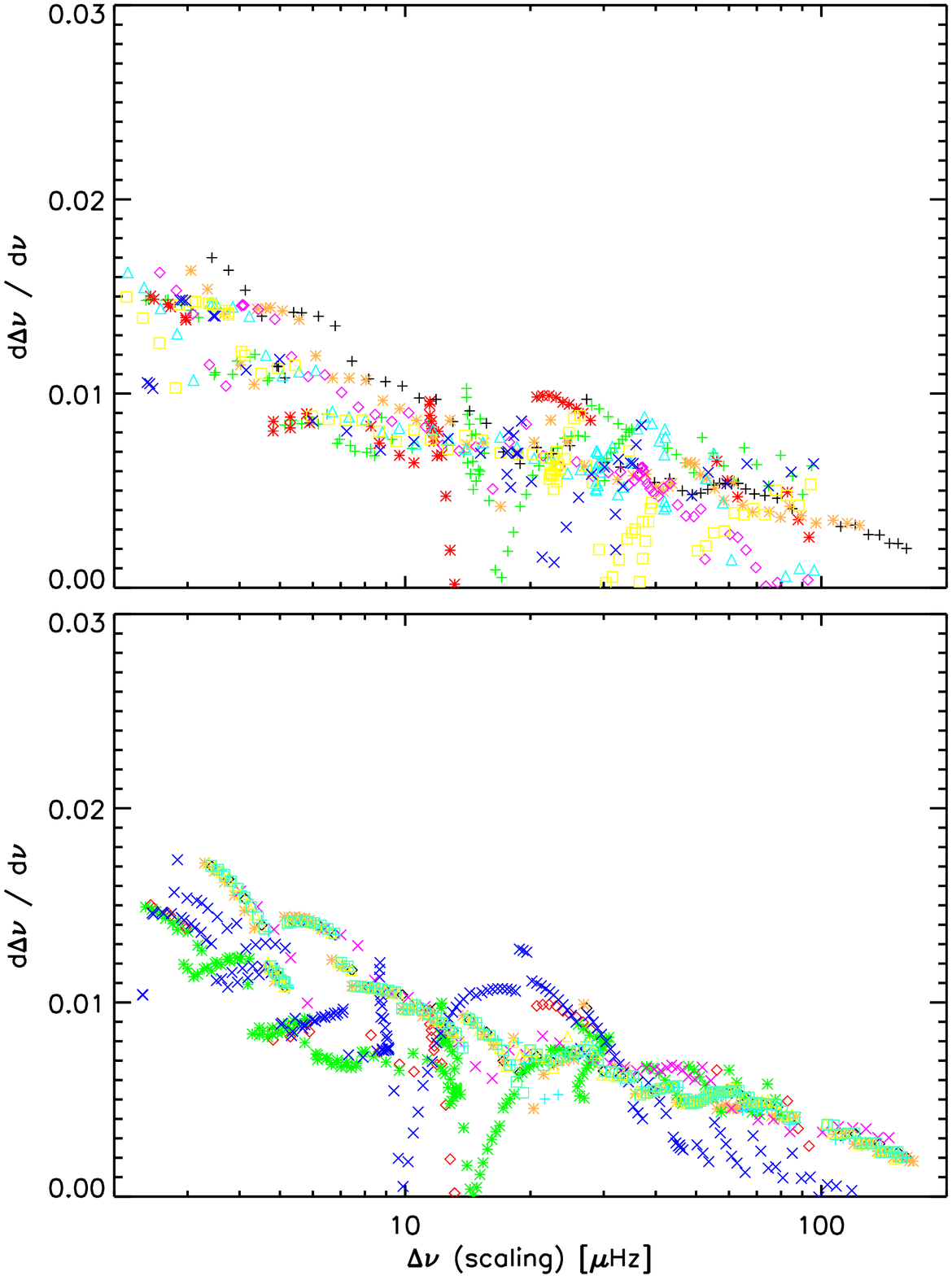}
\end{minipage}
\begin{minipage}{0.45\linewidth}
\centering
\includegraphics[width=\linewidth]{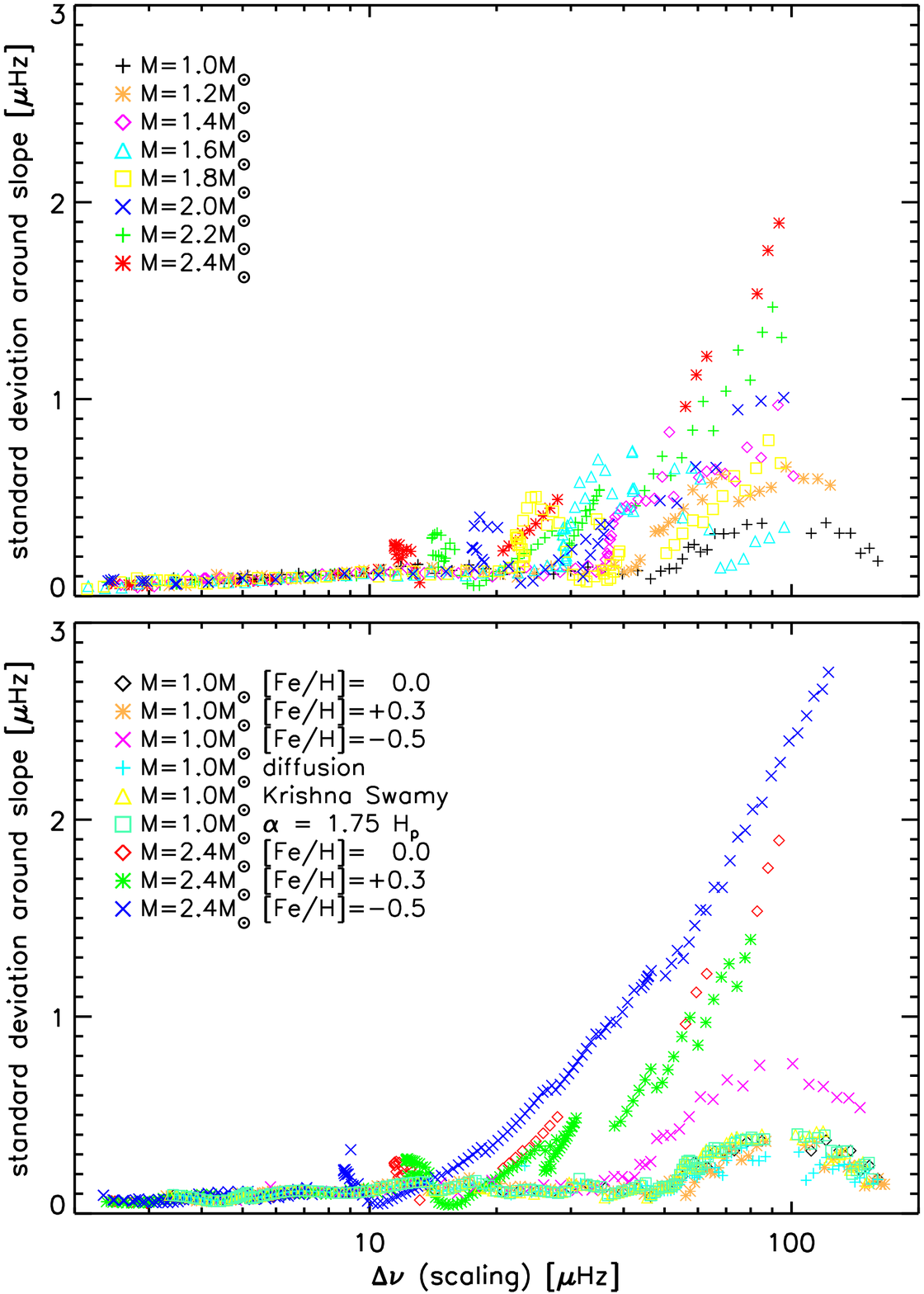}
\end{minipage}
\caption{Left: Linear trend in $\Delta \nu$ over the range $\nu_{\rm max} \pm 4\Delta \nu$ for models with different masses (top) and different metallicity, diffusion, atmosphere or mixing-length parameter (bottom) as a function of $\Delta \nu$ derived from scaling relations. Right: Standard deviation of $\Delta \nu$ after subtracting the linear fit over the range $\nu_{\rm max} \pm 4\Delta \nu$.}
\label{stddev}
\end{figure*}

The ratio of the $\meandnu$ values computed over each frequency range gives an indication of the sensitivity of $\meandnu$ to the frequency range. Fig.~\ref{ratiofit} shows this ratio as a function of stellar radius. The scatter in the ratio of $\meandnu$ computed over different intervals symmetric around $\nu_{\rm max}$ decreases with increasing stellar radii and that the ratios converge to 1.0. 

From the ratios of $\meandnu$ we find that the scatter at low radii ($\lesssim 8$~R$_{\odot}$) reduces for masses below roughly 1.5~M$_{\odot}$. On the other hand no significant change in the scatter is present due to different metallicities or due to the inclusion of diffusion, different atmospheres or mixing-length parameters in the models (see bottom panels of Fig.~\ref{ratiofit}). Note that the relative change in the values of $\meandnu$ for stars with radii $\lesssim 8$~R$_{\odot}$ as a function of the frequency range is of the same order as the precision as with which $\meandnu$ can be determined from current state-of-the-art data. 

\citet{hekker2011comp} suggested two possible reasons for the decreased influence of the frequency range on $\meandnu$ for red giants: 1) the trend in $\Delta \nu$ over a typical frequency range is approximately linear, and/or 2) $\Delta \nu$ changes relatively slowly with frequency. Other reasons might be that in red giants fewer modes are observed, and these modes have lower radial orders
than in less evolved stars. We test these suggestions by fitting a linear polynomial through the values of $\Delta \nu$ over a range $\nu_{\rm max} \pm 4(\Delta \nu)_{\rm scaling}$ and investigate the slope of the fit and the standard deviation of the values around the fit. These results are shown in Fig.~\ref{stddev} and show that the slope or linear trend of the variation in $\Delta \nu$ as a function of frequency for stars with lower $\Delta \nu$, i.e., stars with larger radii, is larger than for stars with higher values of $\Delta \nu$, with little sensitivity to the mass, metallicity, diffusion, atmosphere and mixing-length parameter.  For higher mass models there is some variation visible in the trend, which coincides with the `hook' in the H-R diagram (Fig.~\ref{hrd}), i.e., the contraction during the turn-off of the main sequence. The scatter around the fit is however much lower for stars with $\Delta \nu$ below $\sim$10~$\mu$Hz than for stars with larger values of $\Delta \nu$. Thus we indeed conclude that stars with lower $\Delta \nu$ ($\lesssim 10~\mu$Hz), i.e., larger radii ($\gtrsim 8$~R$_{\odot}$), have a predominantly linear dependence on frequency, while for stars with larger $\Delta \nu$ ($\gtrsim 10~\mu$Hz), i.e., smaller radii ($\lesssim 8$~R$_{\odot}$), the trend is shallower and the scatter in $\Delta \nu$ increases. The amount of scatter does not seem to depend critically on the inclusion of diffusion, different atmosphere or mixing-length parameter in the models, while it increases with mass and with decreasing metallicity. 

\section{Discussion}
Acoustic glitches, i.e., regions of sharp-structure variation in the stellar interior, are known to cause a modulation of $\Delta \nu$ with frequency. For the Sun and other main-sequence stars both the Helium second-ionization zone (He II zone) and the base of the convection zone have been shown to contribute to the modulation of the frequencies \citep[e.g.][and references in the introduction]{monteiro2005,verner2006}. Generally, the signal of the base of the convection zone is weaker than the signal from the He II zone, due to its location deeper in the star. Additionally, \citet{miglio2010} already showed that for a red giant only the He II zone caused a measurable modulation. 

The He II zone causes a local depression in the first adiabatic exponent $\gamma_1~=~(\partial \ln p / \partial \ln \rho)_s$, with $p$ pressure, $\rho$ density and $s$ specific entropy. This glitch modulates the frequencies in a sinusoidal manner with a so-called `period' inversely proportional to the acoustic depth ($\tau$), with 
\begin{equation}
\tau=\int_r^R \frac{dr}{c},
\end{equation}
in which $r$ indicates the radius of the glitch, $R$ the radius of the star and $c$ sound speed. The `period' due to the bottom of the convection zone, located at the position where $\nabla T$ becomes larger than $(\nabla T)_{\rm ad}$, has been computed similarly with respect to the closest acoustic boundary \citep{mazumdar2001b}.

\begin{figure}
\begin{minipage}{\linewidth}
\centering
\includegraphics[width=\linewidth]{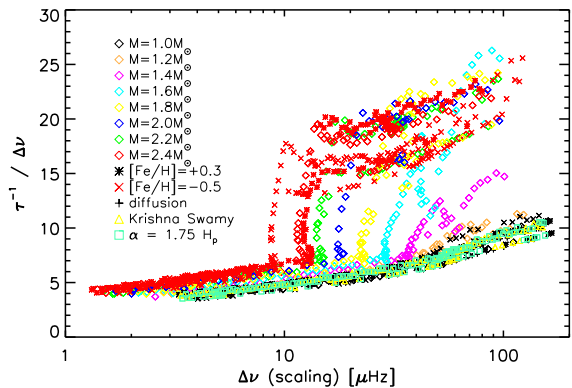}
\end{minipage}
\caption{Period of the frequency modulation expressed in units of $\Delta \nu$ as a function of $\Delta \nu$.}
\label{lambdavsdnu}
\end{figure}

The period due to the He II zone depends on the location of the depression in $\gamma_1$, i.e., a depression close to the surface will cause a lower value for $\tau$ than a depression located deeper in the star. We computed the expected period ($\tau^{-1}$) of the frequency modulation from the location of the He II zone for all our models and divided it by $(\Delta \nu)_{\rm scaling}$ to investigate how many orders one modulation period covers. This is shown in Fig.~\ref{lambdavsdnu}. For models with $\Delta \nu < 10~\mu$Hz the modulation period is about $5 \Delta \nu$ and not significantly dependent on mass, metallicity, diffusion, atmosphere or mixing-length parameter, while the period and the spread in periods increase for models with $\Delta \nu > 10~\mu$Hz.

To determine the dependence of $\meandnu$ on the frequency range we added/removed $2\Delta\nu$ on either side of the frequency interval (Section 3). This is about one full period for models with $\Delta \nu < 10~\mu$Hz and no significant net effect in the mean value of $\Delta \nu$ has been observed. For stars  with $\Delta \nu > 10~\mu$Hz the period depends on mass and is in some cases significantly longer than $5\Delta\nu$ implying that only part of a period is added/removed for the different frequency ranges. This does have an effect on the resulting mean values of $\Delta \nu$. From this we can conclude that for larger stars ($R>8$~R$_{\odot}$) the smooth stellar structure changes reflected in the linear trend of $\Delta\nu$ are dominant over the effect from the He II acoustic glitch, while for smaller stars ($R<8$~R$_{\odot}$) the effect from the He II acoustic glitch is dominant.

\begin{figure}
\begin{minipage}{\linewidth}
\centering
\includegraphics[width=\linewidth]{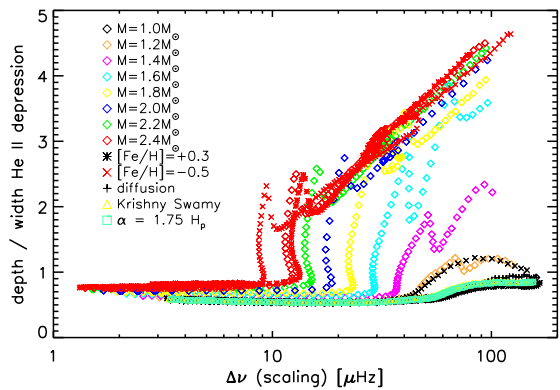}
\end{minipage}
\caption{The depth of the He II depression in $\gamma_1$ divided by the width of the depression expressed in acoustic radius as a function of $\Delta \nu$.}
\label{shape}
\end{figure}

In addition to the period, we also investigated the amplitude of the variations in $\Delta \nu$, parametrized by the standard deviation (see right hand side of Fig.~\ref{stddev}). The amplitude of the modulation is larger for stars with $\Delta \nu \gtrsim 10~\mu$Hz than for stars with $\Delta \nu \lesssim 10~\mu$Hz and depends on mass and metallicity (see Fig.~\ref{stddev}). We expect the amplitude to depend on the shape of the glitch. We parametrize the shape of the glitch by the ratio of the depth of the He II depression in $\gamma_1$, i.e. its strength, to the width (in units of acoustic radius) of the depression, i.e., the narrower the width for a given depth the bigger the effect on the frequencies. The ratios for all models are shown in Fig.~\ref{shape} as a function of $\Delta \nu$.

\begin{figure}
\begin{minipage}{\linewidth}
\centering
\includegraphics[width=\linewidth]{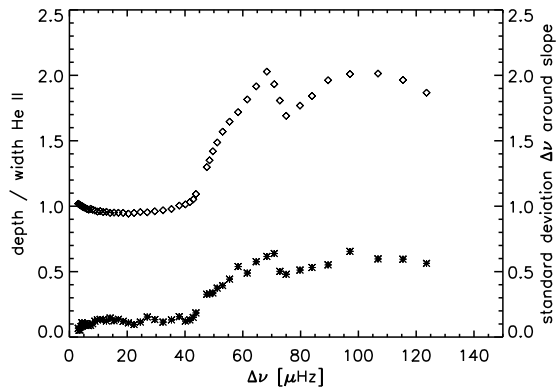}
\end{minipage}
\caption{For $M=1.2$~M$_{\odot}$ models the standard deviation of the variation in $\Delta \nu$ (asterisks) and the depth of the He II depression in $\gamma_1$ divided by the width of the depression expressed in acoustic radius (diamonds) as a function of $\Delta \nu$.}
\label{shapestd}
\end{figure}

The changes in the shape of the He II depression in $\gamma_1$ at a given $\Delta \nu$ are clearly reflected in the changes in the standard deviation of the variation in $\Delta \nu$ at that $\Delta \nu$ (see Fig.~\ref{shapestd}). The general trends in the curves for $\Delta\nu$ and the depth to width ratio of the He II zone are the same, although there are differences in the amplitude of the variations. Therefore we conclude that the shape of the He II depression in $\gamma_1$, i.e. the ratio of the strength over the width of the He II zone, is a significant cause of the difference in the amplitudes of the variation in $\Delta \nu$ which vary with mass and metallicity. However, there must be additional effects that play a role. This could be the base of the convection zone, which induces a small modulation with a period that is of the same order as the period of the He II zone.

We note that the increasing value of $\Delta  \nu$ at which the amplitude of the variation in $\Delta \nu$ increases is a function of  mass and metallicity (Fig.~\ref{stddev}) and that this effect is consistently present in both the period and shape of the He II zone (Figs.~\ref{lambdavsdnu} \& \ref{shape}).

\section{Conclusions}
In this work we confirm the statement by \citet{hekker2011comp} that the value of $\meandnu$ computed for red giants is not as sensitive to the frequency range over which it is computed as is the case for the Sun or other main-sequence stars. The analysis of frequencies from YREC models of stars with different masses and metallicities and with the inclusion of diffusion, a different atmosphere and different mixing-length parameter clearly show that for stars with lower $\Delta \nu$ ($\lesssim 10~\mu$Hz), i.e., larger radii, the dependence of $\Delta \nu$ follows a linear trend with relatively low scatter. For stars with higher $\Delta \nu$ ($\gtrsim 10~\mu$Hz), i.e., smaller radii, the linear trend diminishes, while the scatter increases, with a peak in the scatter at about 100~$\mu$Hz. The magnitude of this peak in the scatter increases with mass and with decreasing metallicity, but does not seem to be sensitive to the inclusion of diffusion, a different atmosphere or mixing-length parameter in the models.

The lower sensitivity of $\meandnu$ on frequency is due to deeper location of a weaker and wider He II zone for extended evolved stars. For less evolved stars the shallower location of a stronger and thinner He II zone causes an increased sensitivity of $\meandnu$ on frequency. The strength and width of the He II zone in less evolved stars changes as a function of mass and metallicity. 

\section*{Acknowledgements}
SH acknowledges financial support from the Netherlands Organisation for Scientific Research (NWO). YE and WJC acknowledge financial support of UK STFC. SH thanks M. Mooij for useful discussions.

\bibliographystyle{mn2e}
\bibliography{dnuI.bib}

\label{lastpage}

\end{document}